\documentclass[aps,pre,reprint,superscriptaddress,longbibliography]{revtex4-2}

\usepackage{amsmath,amssymb,bm}
\usepackage{graphicx}
\usepackage{booktabs}
\usepackage{dcolumn}
\usepackage{hyperref}

\begin{document}

\title{Optimization versus fair sampling in quantum annealing: An exactly characterized frustrated Ising benchmark}

\author{Masaru Hitomi}
\affiliation{Graduate School of Information Sciences, Tohoku University, Sendai 980-8579, Japan}

\author{Masayuki Ohzeki}
\affiliation{Graduate School of Information Sciences, Tohoku University, Sendai 980-8579, Japan}
\affiliation{Department of Physics, Institute of Science Tokyo, Tokyo 152-8551, Japan}
\affiliation{Research and Education Institute for Semiconductors and Informatics, Kumamoto University, Kumamoto 860-0862, Japan}
\affiliation{Sigma-i Co., Ltd., Tokyo 108-0075, Japan}

\date{\today}

\begin{abstract}
We show that reaching a degenerate ground-state manifold and sampling it uniformly are distinct properties of a quantum annealer, using an Ising ring with additional couplings between maximally separated sites whose exact ground-state manifold is known for every system size.
The interaction graph depends on the parity of the system size $N$.
For antiferromagnetic couplings the zero-field ground-state manifold is classified in closed form for each congruence class of $N$ modulo four, including its exact degeneracy and its decomposition into defect sectors.
Every sampled configuration can therefore be assigned unambiguously to the ground-state manifold or to an excited state, which separates the efficiency of reaching the manifold from the distribution realized within it.
We embed instances with up to $162$ logical spins on a D-Wave Advantage processor using length-two chains, and compare standard and fast annealing with simulated annealing applied to the embedded and to the logical problem.
For a frustrated odd instance the processor reaches the ground-state manifold far more often than the classical baselines and visits $164$ of the $166$ exact ground states, yet the conditional distribution stays clearly nonuniform over both ground states and defect sectors.
The two properties need not improve together: for $N=160$ the processor attains the larger ground-state probability ($0.869$ against $0.519$) but also the larger total-variation distance from uniformity ($0.317$ against $0.136$).
A pause in the intermediate part of the anneal raises the ground-state probability from $0.42$ to $0.86$ while leaving the residual bias unchanged, locating a relaxation window where transverse-field fluctuations and environmental relaxation act together and supporting a thermally assisted quantum-annealing picture.
Reaching a degenerate ground-state manifold and sampling it uniformly are therefore distinct properties of an annealing protocol.

\end{abstract}

\maketitle

\section{Introduction}
Quantum annealing
\cite{kadowaki1998quantum, farhi2001quantum, johnson2011quantum, albash2018adiabatic, hauke2020perspectives} 
is commonly assessed by its ability to find low-energy or ground-state configurations of an optimization problem
\cite{ronnow2014defining, boixo2014evidence, denchev2016what}. 
For problems with a degenerate ground-state manifold, however, reaching the ground-state energy is only one aspect of the computation. 
Even when an annealer reaches the ground-state manifold with high probability, 
it may sample only a biased subset of the exactly degenerate solutions
\cite{matsuda2009ground, mandra2017exponentially, zhang2017advantages, konz2019uncertain, king2016degeneracy}. 
Optimization success and sampling quality are therefore distinct properties of an annealing process, 
and characterizing both is important for understanding how an annealer explores a complex energy landscape
\cite{douglass2015constructing, amin2018quantum}.

This distinction is difficult to quantify in generic optimization problems. 
If the complete set of ground states is not known independently, an observed nonuniform output distribution can reflect the structure of the solution space itself~\cite{king2016degeneracy}, finite-sampling effects, embedding details~\cite{choi2008minor, venturelli2015quantum}, or device-dependent biases~\cite{perdomoortiz2016determination, kadowaki2019experimental}. 
A useful benchmark for this purpose should therefore provide more than a known minimum energy: 
the complete ground-state manifold should be characterized exactly. 
Such a setting makes it possible to ask separately how efficiently an annealer reaches the ground-state manifold and how it samples the manifold after reaching it. 
This second question is the fair-sampling problem
\cite{matsuda2009ground, mandra2017exponentially, zhu2019fair, yamamoto2020fair, kumar2020achieving, pelofske2021sampling}, 
which is invisible if the performance of an annealer is characterized only by the best energy or the ground-state success probability.
Existing benchmarks meet these requirements only partially. 
Problems with planted solutions~\cite{hen2015probing, wang2017patch, perera2020computational} fix the ground-state energy but leave the size and structure of the solution set unknown, 
so that no reference distribution is available. 
Disordered models such as spin glasses~\cite{katzgraber2014glassy, katzgraber2015seeking} require the ground-state manifold to be determined numerically, and its completeness cannot be guaranteed at the sizes of interest~\cite{zhu2019fair, ochoa2019feeding}. 
Small frustrated units~\cite{matsuda2009ground}, for which the degenerate states can be enumerated directly, do provide a well-defined reference, but they involve only a few spins and therefore do not reach the scale at which embedding and freeze-out effects become relevant. 
What is required instead is a family of problems whose complete ground-state manifold is known in closed form at arbitrary system size.

In this paper,
we use an Ising ring with additional interactions between maximally separated sites as such a benchmark. 
Although the microscopic interaction rule is the same for all system sizes, the interaction graph depends on the parity of $N$. 
For even $N$, 
each spin has a unique antipodal partner and the graph is a M\"obius ladder~\cite{guy1967mobius}, 
whereas for odd $N$, 
each spin has two maximally separated neighbors and the graph is the circulant graph $C_N(1,(N-1)/2)$~\cite{boesch1984circulants}. 
The odd-$N$ graph can equivalently be relabeled as $C_N(1,2)$. 
This parity dependence generates distinct frustration patterns
\cite{wannier1950antiferromagnetism, toulouse1977theory}, 
and in the antiferromagnetic regime considered here the zero-field ground-state manifold depends qualitatively on $N$ modulo four. 
Importantly for the present purpose, its energy, degeneracy, and defect structure can be classified analytically.

This exact ground-state classification allows every decoded sample to be assigned unambiguously either to the ground-state manifold or to an excited state. 
It also allows us to distinguish two quantitative questions. 
The first is optimization: 
how often does the annealer reach the exact ground-state manifold? We characterize this by the ground-state hit probability $P_G$~\cite{ronnow2014defining, boixo2014evidence}. 
The second is sampling: 
conditional on reaching the ground-state manifold, how close is the resulting distribution to the uniform distribution over the exactly degenerate ground states? 
In the frustrated instances, the ground states can furthermore be organized into defect sectors by identifying global-spin-flip partners,
allowing the sampling bias to be examined both at the level of individual states and at the level of the underlying defect structure.

We implement the model on a D-Wave quantum annealer~\cite{johnson2011quantum, boothby2020next} using manually constructed embeddings~\cite{choi2008minor, cai2014practical} in which all logical spins are represented by short chains~\cite{vinci2015quantum}. 
This controlled embedding~\cite{marshall2020perils} allows us to compare the quantum-annealing results with simulated annealing~\cite{kirkpatrick1983optimization, isakov2015optimized} both on the embedded physical problem and directly on the logical problem. 
We also compare fast and standard annealing schedules in order to probe how the ability to reach the exact ground-state manifold changes with the annealing time.

For the frustrated odd-$N$ instance studied in detail, 
the quantum processor reaches the analytically known ground-state manifold substantially more frequently than the simulated-annealing baselines used here. 
At the same time, 
the conditional distribution over the exact ground states remains clearly nonuniform. 
Thus, successful optimization does not imply fair sampling~\cite{matsuda2009ground, mandra2017exponentially, konz2019uncertain}: 
the processor can efficiently reach the correct energy manifold while retaining a measurable preference for particular ground states and defect sectors. 
This separation is also observed in other congruence classes, showing that the ground-state hit probability alone does not fully characterize the behavior of the annealer.

We further probe the dynamics by introducing a pause during the annealing schedule~\cite{marshall2019power, chen2020why, passarelli2019improving, gonzalezizquierdo2021ferromagnetically}. 
A pause in an intermediate region of the anneal produces a pronounced enhancement of the ground-state hit probability, indicating efficient relaxation~\cite{marshall2019power, marshall2017thermalization} into the ground-state manifold while the transverse field remains appreciable. 
The enhancement is strongly dependent on the pause location and disappears when the pause is introduced too early or too late~\cite{marshall2019power, chen2020why}. 
At the same time, the conditional distribution within the degenerate ground-state manifold remains nonuniform even where the ground-state probability is strongly enhanced. 
The combined annealing-time and pause dependence therefore supports a thermally assisted quantum-annealing picture
\cite{dickson2013thermally, amin2009role, albash2012quantum, arceci2017dissipative, smelyanskiy2017quantum}, 
in which transverse-field fluctuations and environmental relaxation cooperate in driving the system toward low-energy configurations, while freeze-out
\cite{amin2015searching, marshall2017thermalization}, 
analog biases~\cite{perdomoortiz2016determination, pearson2019analog, zhu2016best}, 
and the embedding geometry~\cite{marshall2020perils, choi2008minor}
 can leave a residual sampling bias within the ground-state manifold.

The maximally separated Ising ring thus provides a compact and exactly characterized benchmark for separating optimization success from fair sampling in open-system quantum annealing. 
More broadly, our results show that relaxation into a degenerate ground-state manifold and redistribution within that manifold need not occur in the same way. 
Characterizing both processes is therefore essential for understanding the output of a quantum annealer beyond conventional ground-state success metrics.

The paper is organized as follows.
Section~\ref{sec_model} resents the exactly characterized benchmark model and its ground-state classification.
We next describe the annealing implementation and the sampling diagnostics in Section~\ref{sec_met}.
Section~\ref{sec_res} presents the results and 
Sec.~\ref{sec_dis} discusses the implications of these findings, and 
we summarize the paper in Sec.~\ref{sec_con}.

\begin{figure}[t]
    \centering
    \includegraphics[width=0.9\columnwidth]{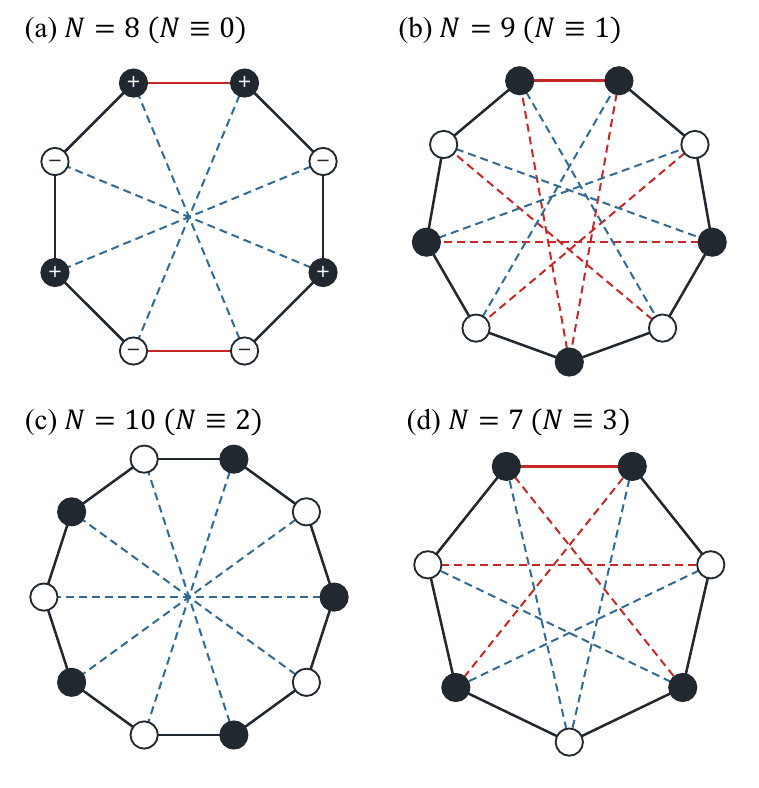}
    \caption{
      Schematic picture of representative ground states for the couplings in Eq.~\eqref{eq:benchmark_couplings}.
      Filled and open circles denote spins $+1$ and $-1$.
      Solid and dashed lines represent nearest-neighbor and maximally
      separated bonds, respectively. Red marks unsatisfied bonds.
      (a) $N=8$: two antipodal nearest-neighbor defects, with all
      maximally separated bonds satisfied.
      (b),(d) $N=9,7$: one nearest-neighbor defect, with five and three
      unsatisfied maximally separated bonds, respectively.
      (c) $N=10$: a N\'eel state satisfying all bonds.
    }
    \label{fig:model}
\end{figure}

\section{Exactly characterized benchmark model}
\label{sec_model}

\subsection{Model}
\label{sub_model}

We consider $N$ Ising spins $\sigma_i=\pm1$ on a ring, with a nearest-neighbor interaction and an additional interaction between maximally separated sites. 
At zero longitudinal field, the Hamiltonian is written as
\begin{align}
  H_N(\boldsymbol{\sigma})
  =
  -J_1 \sum_{(i,j)\in E_1} \sigma_i \sigma_j
  -J_* \sum_{(i,j)\in E_*} \sigma_i \sigma_j ,
  \label{eq:hamiltonian}
\end{align}
where $E_1$ denotes the nearest-neighbor edge set and $E_*$ denotes the set of maximally separated edges. 
Site indices are understood modulo $N$.

Although Eq.~\eqref{eq:hamiltonian} is defined by the same microscopic rule for all $N$, 
the interaction graph depends on the parity of the system size. 
For even $N$, each site has a unique antipodal partner, and
\begin{align}
  E_*
  =
  \left\{
  (i,i+N/2)
  \,\middle|\,
  i=0,\ldots,N/2-1
  \right\}.
  \label{eq:estar_even}
\end{align}
The resulting graph is a M\"obius ladder, consisting of the original ring together with a perfect matching between antipodal sites~\cite{guy1967mobius, hamerly2019experimental, kalinin2022computational, cummins2025ising}.

For odd $N=2m+1$, 
there is no unique antipodal site. 
Instead, each site has two maximally separated neighbors, and the maximally separated edges can be written as
\begin{align}
  m &= \frac{N-1}{2},
  \\
  E_*
  &=
  \left\{
  (i,i+m)
  \,\middle|\,
  i=0,\ldots,N-1
  \right\}.
  \label{eq:estar_odd}
\end{align}
The interaction graph is therefore the circulant graph $C_N(1,m)$~\cite{boesch1984circulants}. 
Since $\gcd(m,N)=1$, a relabeling of the sites by
\begin{align}
  \hat{\sigma}_j
  =
  \sigma_{jm \bmod N}
  \label{eq:odd_relabeling}
\end{align}
maps the maximally separated edges to step-one edges. 
Using $2m=N-1\equiv -1 \pmod N$, the original nearest-neighbor edges become step-two edges. 
The odd-$N$ Hamiltonian can thus equivalently be written as
\begin{align}
  H_N
  =
  -J_* \sum_{j=0}^{N-1}
  \hat{\sigma}_j \hat{\sigma}_{j+1}
  -J_1 \sum_{j=0}^{N-1}
  \hat{\sigma}_j \hat{\sigma}_{j+2}.
  \label{eq:odd_relabelled_hamiltonian}
\end{align}
This representation is useful below because it makes the frustrated cycle structure of the odd system explicit.

Throughout the annealing experiments, we restrict attention to the antiferromagnetic benchmark
\begin{align}
  J_1 &= -1,
  \nonumber\\
  J_* &=
  \begin{cases}
  -1, & N\ {\rm even},\\[2pt]
  -\dfrac{1}{2}, & N\ {\rm odd}.
  \end{cases}
  \label{eq:benchmark_couplings}
\end{align}
There is no longitudinal field. 
The parity-dependent normalization of $J_*$ compensates for the different density of maximally separated bonds. 
In fact,
\begin{align}
  |E_*|
  =
  \begin{cases}
  N/2, & N\ {\rm even},\\
  N,   & N\ {\rm odd},
  \end{cases}
  \label{eq:estar_number}
\end{align}
so that each site is incident on one maximally separated bond for even $N$ and two for odd $N$. 
With the choice in Eq.~\eqref{eq:benchmark_couplings}, the total absolute coupling weight associated with the maximally separated interactions is the same,
$|J_*||E_*|=N/2$, for the two parity sectors. 
This makes the comparison between even and odd instances less sensitive to the trivial difference in bond density while preserving their distinct graph topology and frustration structure.

In the D-Wave Ising convention used below, 
the corresponding problem couplers $-J_1$ and $-J_*$ are positive and therefore antiferromagnetic.

\subsection{Exact ground-state structure}
\label{sub_gs}

A central feature of the benchmark is that its zero-field ground-state manifold can be classified analytically. 
The classification depends on $N$ modulo four because the antiferromagnetic ordering preferred along the ring must also satisfy the parity-dependent maximally separated interactions. 
Here we summarize only the results required for the annealing analysis and a derivation in terms of domain-wall and parity constraints is given in Appendix~\ref{app_gs}.
Figure~\ref{fig:model} illustrates the interaction graphs and
representative ground states for the four classes of $N$ modulo four.

For even $N$, 
the distinction arises from the orientation of antipodal spins in a N\'eel configuration. 
When $N\equiv2\pmod4$, an alternating N\'eel state makes every nearest-neighbor bond antiparallel and also makes every antipodal pair antiparallel. 
All bonds are therefore simultaneously satisfied, and the only ground states are the two globally spin-reversed N\'eel configurations. 
For the benchmark couplings,
\begin{align}
  N &\equiv 2 \pmod 4:
  &
  E_0 &= -\frac{3N}{2},
  &
  |G| &= 2.
  \label{eq:gs_even2}
\end{align}

For $N\equiv0\pmod4$, 
the alternating configuration instead places antipodal spins in the same orientation. 
The antipodal antiferromagnetic interaction is therefore incompatible with a defect-free N\'eel pattern. 
The minimum-energy configurations resolve this frustration by introducing two nearest-neighbor defects while satisfying the maximally separated bonds~\cite{cummins2025ising, kalinin2022computational}. 
Their positions generate an $N$-fold degenerate ground-state manifold,
\begin{align}
  N &\equiv 0 \pmod 4:
  &
  E_0 &= -\frac{3N}{2}+4,
  &
  |G| &= N.
  \label{eq:gs_even0}
\end{align}
Thus the even subsequence already separates into an unfrustrated two-state manifold for $N\equiv2\pmod4$ and an extensively position-labeled frustrated manifold for $N\equiv0\pmod4$.

For odd $N$, 
the relabeled step-two edges in Eq.~\eqref{eq:odd_relabelled_hamiltonian} form an odd cycle. 
An antiferromagnetic arrangement on this cycle necessarily contains one unavoidable defect. 
Once the defect position is specified, the spin configuration is fixed up to a global spin reversal. 
The exact ground-state manifold therefore contains
\begin{align}
  |G| = 2N
  \label{eq:gs_odd_degeneracy}
\end{align}
states. 
The contribution from the $J_*$ bonds depends on $N$ modulo four, giving
\begin{align}
  E_0
  =
  \begin{cases}
  -N+\dfrac{5}{2},
  & N\equiv1\pmod4,
  \\[6pt]
  -N+\dfrac{3}{2},
  & N\equiv3\pmod4.
  \end{cases}
  \label{eq:gs_odd_energy}
\end{align}
The two states related by a global spin flip have the same defect position. 
It is therefore useful to group them into a single defect sector. 
Denoting the set of such sectors by $S$, one has
\begin{align}
  |S| = N,
  \qquad
  |G| = 2|S| = 2N.
  \label{eq:defect_sectors}
\end{align}
This sector decomposition will be used below to distinguish a bias in the spatial distribution of defects from an imbalance between globally spin-reversed partners.

The ground-state structure relevant to the benchmark is summarized in Table~\ref{tab:ground_state_summary}. 
The important point for the annealing analysis is that both the complete ground-state set and its defect-sector organization are known in advance. 
Consequently, 
each decoded sample can be classified without an optimization heuristic, and the probability of reaching the exact ground-state manifold can be separated from the conditional distribution within that manifold.

\begin{table}[t]
  \caption{
  Exact zero-field ground-state structure for the antiferromagnetic
  benchmark of Eq.~\eqref{eq:benchmark_couplings}.  A defect denotes an
  unsatisfied nearest-neighbor bond in the ordering representation
  appropriate to each parity sector.  For odd $N$, the $2N$ ground
  states form $N$ defect sectors related pairwise by global spin
  reversal.
  }
  \label{tab:ground_state_summary}
  \centering
  \begin{tabular}{c c c c}
  \hline\hline
  $N \bmod 4$ &
  ground-state structure &
  $E_0$ &
  $|G|$
  \\
  \hline
  $0$ &
  two-defect frustrated states &
  $-3N/2+4$ &
  $N$
  \\
  $1$ &
  one-defect odd-cycle states &
  $-N+5/2$ &
  $2N$
  \\
  $2$ &
  N\'eel pair &
  $-3N/2$ &
  $2$
  \\
  $3$ &
  one-defect odd-cycle states &
  $-N+3/2$ &
  $2N$
  \\
  \hline\hline
  \end{tabular}
\end{table}

For later reference, the frustrated odd instance emphasized in the sampling analysis belongs to the $N\equiv3\pmod4$ class. 
Its exact ground-state manifold and defect sectors are therefore obtained directly from Eqs.~\eqref{eq:gs_odd_degeneracy}--\eqref{eq:defect_sectors};
for $N=83$, this gives
\begin{align}
  |G| &= 166,
  &
  |S| &= 83,
  &
  E_0 &= -81.5.
  \label{eq:n83_groundstate}
\end{align}

\section{Annealing implementation and sampling diagnostics}
\label{sec_met}

\subsection{Annealing implementation and protocols}
\label{sub_ann}

We manually embed the logical Ising problems on a D-Wave Advantage processor~\cite{boothby2020next, mcgeoch2022advantage} using chains of length two for all logical spins. 
The ferromagnetic chain coupling is set to
\begin{align}
  J_{\rm chain}=-2.
  \label{eq:chain_coupling}
\end{align}
Decoded logical spins are obtained by majority vote. 
For the representative $N=83$ instance, the embedding uses $83$ logical spins and $166$ physical qubits.

The experiments were performed on D-wave Advantage system4. 
The exported working graph contained $5627$ active qubits and $40279$ active couplers.
All QPU calls used \texttt{auto\_scale=True}. 
To compare samples on the same footing, all reported energies were recomputed after decoding using the original logical Hamiltonian in Eq.~\eqref{eq:hamiltonian}.

We compare the QPU results with two simulated-annealing baselines~\cite{oshiyama2022benchmark}.
Physical simulated annealing (physical SA) is applied directly to the embedded physical Ising problem, followed by the same decoding procedure as for the QPU samples. 
Logical simulated annealing (logical SA) is instead applied directly to the logical Ising problem, without chain variables. 
Both SA protocols use $3000$ reads and $1000$ sweeps per read, while each QPU data set contains $3000$ reads.

To examine the annealing-time dependence, 
we use fast anneals with annealing times
\begin{align}
  t_a = 0.005,\ 0.05,\ 0.1~\mu{\rm s},
\end{align}
and standard anneals with
\begin{align}
  t_a = 0.5,\ 2,\ 20~\mu{\rm s}.
\end{align}

We also perform pause annealing for the frustrated odd-$N$ instance studied in detail. 
Taking the standard $20~\mu{\rm s}$ anneal as the baseline, we insert a $100~\mu{\rm s}$ pause at annealing fraction $s_p$ using the schedule
\begin{align}
  (0,0)
  \rightarrow
  (20s_p,s_p)
  \rightarrow
  (20s_p+100~\mu{\rm s},s_p)
  \rightarrow
  (120~\mu{\rm s},1).
  \label{eq:pause_schedule}
\end{align}
We first scan $s_p=0.2,0.3,\ldots,0.9$ and then resolve the region of strongest relaxation using $s_p=0.34,0.36,\ldots,0.46$.

\subsection{Ground-state sampling diagnostics}
\label{sub_samp}

Because the exact ground-state manifold $G$ is known analytically, each decoded sample can be classified unambiguously as either a ground state or an excited state. 
We quantify optimization performance by the ground-state hit probability
\begin{align}
  P_G
  =
  \frac{
  \#\{\text{decoded ground-state samples}\}
  }{
  \#\{\text{all decoded samples}\}
  }.
  \label{eq:ground_state_probability}
\end{align}

To characterize sampling within the ground-state manifold, let $p_\alpha$ denote the empirical probability of ground state $\alpha\in G$, conditioned on the decoded sample belonging to $G$.
We compare $p_\alpha$ with the uniform distribution
\begin{align}
  u_\alpha=\frac{1}{|G|}.
\end{align}
The primary measure of sampling bias is the total-variation distance,
\begin{align}
  d_{\rm TV}(p,u)
  =
  \frac{1}{2}
  \sum_{\alpha\in G}
  \left|p_\alpha-u_\alpha\right|.
  \label{eq:tv_distance}
\end{align}

For the frustrated manifolds, we additionally group globally spin-reversed ground states into the defect sectors defined in Sec.~\ref{sub_gs}. 
The same diagnostics are then applied to the corresponding sector distribution. 
We distinguish state-level and sector-level total-variation distances by $d_{\rm TV}^{\rm st}$ and $d_{\rm TV}^{\rm sec}$, respectively.
The former is sensitive to all nonuniformity within the ground-state manifold, whereas the latter directly probes bias among distinct defect positions.

\section{Results}
\label{sec_res}

\subsection{Annealing-time dependence}
\label{sec_res_ann}

\begin{figure*}[t]
    \centering
    \includegraphics[width=\textwidth]{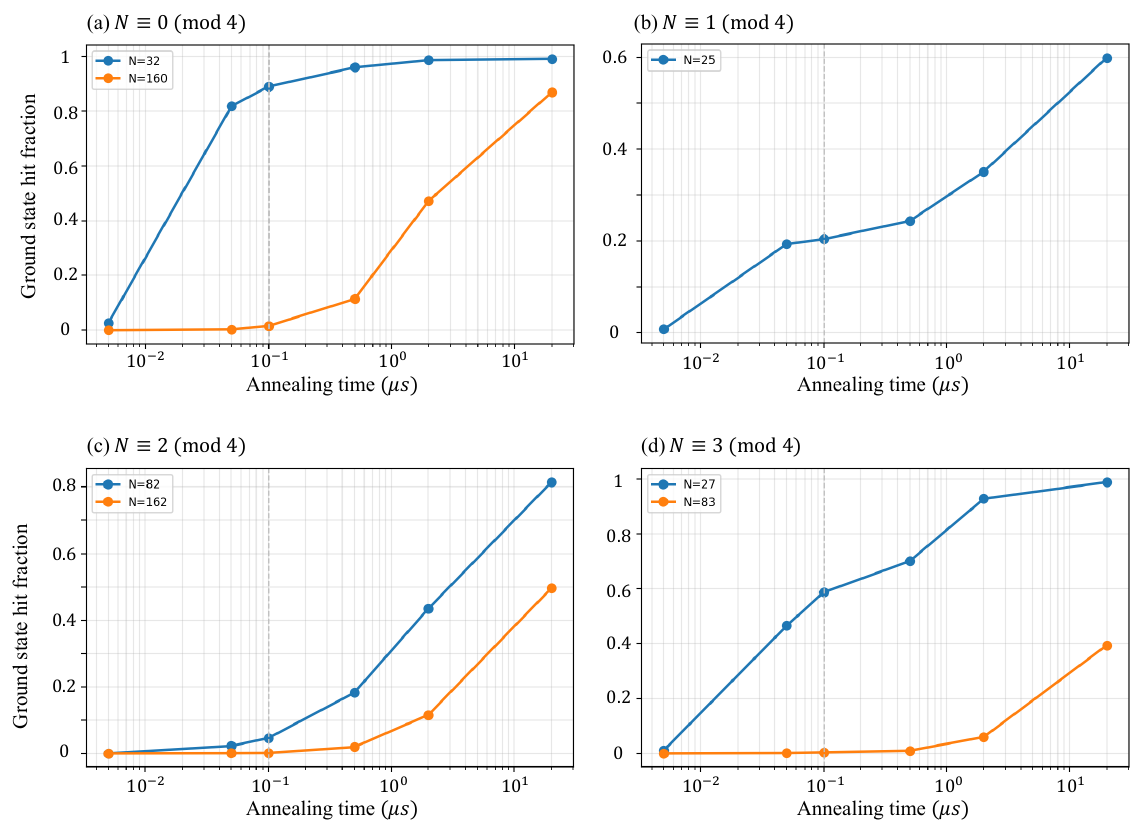}
    \caption{
    Ground-state hit probability $P_G$ of the QPU as a function of the
    annealing time for representative system sizes in the four
    congruence classes of $N$ modulo four:
    (a) $N\equiv0$, (b) $N\equiv1$, (c) $N\equiv2$, and
    (d) $N\equiv3 \pmod 4$.
    The vertical dashed line at $0.1~\mu{\rm s}$ marks the upper end of the fast-anneal range used here; the longer-time points correspond to standard annealing.
    The larger instances exhibit a pronounced increase in $P_G$ with increasing standard annealing time.
    }
    \label{fig:annealing_time}
\end{figure*}

Figure~\ref{fig:annealing_time} shows the ground-state hit probability $P_G$ of the QPU as a function of the annealing time for representative system sizes in all four congruence classes of $N$ modulo four.
The dependence on annealing time becomes particularly pronounced for the larger instances. 
In the fast-anneal regime, their ground-state hit probabilities remain small, whereas standard annealing produces a substantial increase in $P_G$.

For example, for the frustrated odd instance $N=83$, the ground-state hit probability remains close to zero in the fast-anneal regime and increases to $P_G=0.394$ for the standard $20~\mu{\rm s}$ anneal.
For the larger even instances, the corresponding $20~\mu{\rm s}$ values are $P_G=0.869$ for $N=160$ and $P_G=0.496$ for $N=162$.
The systematic improvement with increasing standard annealing time shows that the final ground-state population is strongly dependent on the available annealing time rather than being determined by the short-time dynamics alone.

\subsection{Optimization and fair sampling are distinct}
\label{sec:optimization_sampling}

We next compare optimization performance with sampling within the exactly degenerate ground-state manifold.
Table~\ref{tab:fair_sampling} gives representative results for physical SA, logical SA, and the QPU with a standard $20~\mu{\rm s}$ anneal.
The frustrated $N=83$ instance provides the clearest example.
Physical SA produces only $21$ ground-state samples in $3000$ reads, corresponding to $P_G=0.007$, 
while logical SA gives $P_G=0.023$ and observes $60$ of the $166$ exact ground states.
The large values of $d_{\rm TV}^{\rm st}$ and $d_{\rm TV}^{\rm sec}$ for these SA data should therefore be interpreted with care, 
the ground-state manifold is sampled too sparsely for its full support to be represented.

In contrast, 
the QPU reaches the $N=83$ ground-state manifold with $P_G=0.394$ and observes $164$ of the $166$ exact ground states.
This substantially larger optimization success does not, however, produce a uniform conditional distribution within the manifold.
The QPU data give $d_{\rm TV}^{\rm st}=0.300$ at the state level and $d_{\rm TV}^{\rm sec}=0.165$ at the defect-sector level.
Thus, reaching the exact ground-state manifold and sampling that manifold uniformly constitute distinct criteria.

\begin{table*}[t]
\caption{
Representative ground-state optimization and sampling diagnostics.
$P_G$ is the decoded ground-state hit probability.
$d_{\rm TV}^{\rm st}$ and $d_{\rm TV}^{\rm sec}$ are the
total-variation distances from the uniform distributions over exact
ground states and defect sectors, respectively.
The last column gives the number of observed exact ground states
relative to the exact ground-state degeneracy $|G|$.
}
\label{tab:fair_sampling}
\centering
\begin{ruledtabular}
\begin{tabular}{ccccccc}
$N$ &
$N\bmod4$ &
protocol &
$P_G$ &
$d_{\rm TV}^{\rm st}$ &
$d_{\rm TV}^{\rm sec}$ &
support$/|G|$
\\
\hline
83  & 3 & physical SA       & 0.007 & 0.873 & 0.759 & 21/166  \\
83  & 3 & logical SA        & 0.023 & 0.639 & 0.434 & 60/166  \\
83  & 3 & QPU, $20~\mu{\rm s}$ & 0.394 & 0.300 & 0.165 & 164/166 \\
160 & 0 & physical SA       & 0.066 & 0.365 & 0.254 & 114/160 \\
160 & 0 & logical SA        & 0.519 & 0.136 & 0.089 & 160/160 \\
160 & 0 & QPU, $20~\mu{\rm s}$ & 0.869 & 0.317 & 0.182 & 160/160 \\
162 & 2 & logical SA        & 0.149 & 0.009 & 0.000 & 2/2     \\
162 & 2 & QPU, $20~\mu{\rm s}$ & 0.496 & 0.419 & 0.000 & 2/2     \\
\end{tabular}
\end{ruledtabular}
\end{table*}

\begin{figure}[t]
    \centering
    \includegraphics[width=\columnwidth]{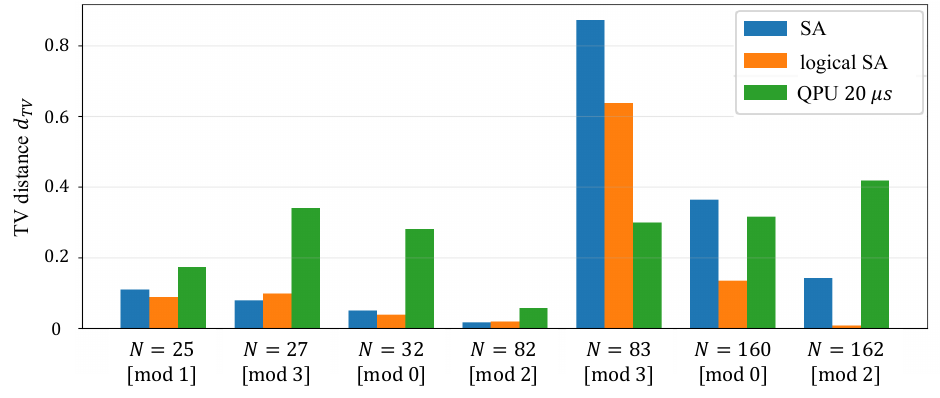}
    \caption{
    State-level total-variation distance
    $d_{\rm TV}^{\rm st}$ from the uniform distribution over exact
    ground states for representative instances.
    ``SA'' denotes simulated annealing on the embedded physical
    problem, ``logical SA'' denotes simulated annealing directly on
    the logical problem, and the QPU results use a standard
    $20~\mu{\rm s}$ anneal.
    A smaller value indicates a distribution closer to uniformity.
    The comparison shows that a protocol with a larger ground-state
    hit probability need not sample the degenerate ground-state
    manifold more uniformly.
    }
    \label{fig:fair_sampling_tv}
\end{figure}

Figure~\ref{fig:fair_sampling_tv} extends the comparison of $d_{\rm TV}^{\rm st}$ over representative instances from all four congruence classes.
There is no general correspondence between a large ground-state hit probability and a small total-variation distance.
For example, for $N=160$, logical SA gives $d_{\rm TV}^{\rm st}=0.136$, whereas the QPU gives $d_{\rm TV}^{\rm st}=0.317$, 
even though the QPU has the substantially larger ground-state hit probability ($P_G=0.869$ compared with $0.519$).

The $N=162$ case gives an especially simple illustration because the ground-state manifold consists only of the two globally spin-reversed N\'eel states.
Logical SA samples the two states nearly uniformly, $d_{\rm TV}^{\rm st}=0.009$, whereas the QPU output at $20~\mu{\rm s}$ is strongly asymmetric.
The observed probabilities of the two ground states are approximately $0.919$ and $0.081$, yielding $d_{\rm TV}^{\rm st}\simeq0.419$.
After the two global-spin-flip partners are combined into a single sector, the sector-level distance is trivially zero, as reflected in Table~\ref{tab:fair_sampling}.
These results demonstrate that optimization success and sampling uniformity need not improve together.

\subsection{Pause-assisted relaxation and residual sampling bias}
\label{sub_pause}

\begin{figure*}[t]
    \centering
    \includegraphics[width=\textwidth]{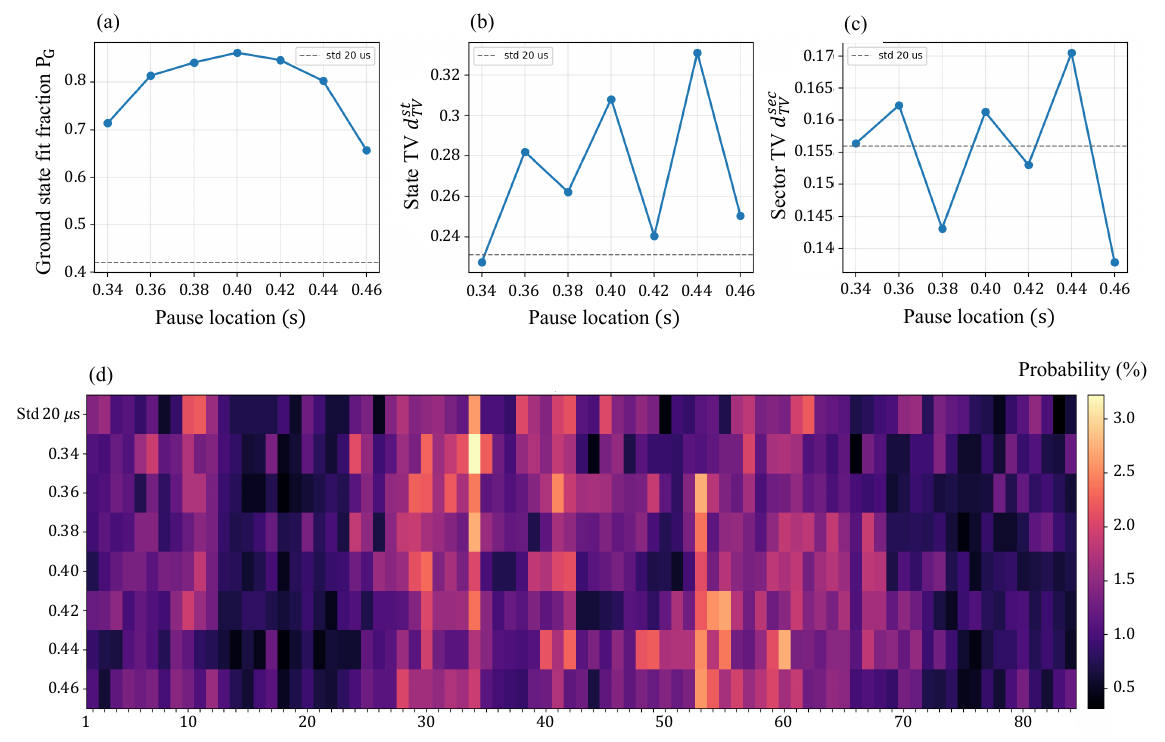}
    \caption{
    Pause-annealing results for $N=83$ with a
    $100~\mu{\rm s}$ pause.
    (a) Ground-state hit probability $P_G$,
    (b) state-level total-variation distance
    $d_{\rm TV}^{\rm st}$, and
    (c) sector-level total-variation distance
    $d_{\rm TV}^{\rm sec}$ as functions of the pause location $s_p$.
    The horizontal dashed lines denote the corresponding values for
    the standard $20~\mu{\rm s}$ baseline acquired with the pause
    data.
    (d) Conditional probabilities of the $83$ defect sectors for the
    standard anneal and the refined pause locations; each row is
    normalized over the exact ground-state samples in that data set.
    The pause produces a pronounced enhancement of $P_G$ near
    $s_p\simeq0.4$, while the conditional state and defect-sector
    distributions remain nonuniform.
    }
    \label{fig:pause_summary}
\end{figure*}

The $N=83$ instance is useful for probing the annealing dynamics because the standard QPU anneal reaches the ground-state manifold much more frequently than the SA baselines while retaining a clearly nonuniform distribution within that manifold.
Figure~\ref{fig:pause_summary} summarizes the refined pause scan.
The standard $20~\mu{\rm s}$ baseline acquired together with the pause data gives $P_G=0.422$.
This baseline is a separate data set from the $20~\mu{\rm s}$ QPU sample reported in Table~\ref{tab:fair_sampling}, for which $P_G=0.394$.

Introducing a $100~\mu{\rm s}$ pause produces a broad enhancement of the ground-state population for approximately $0.36\lesssim s_p\lesssim0.44$ [Fig.~\ref{fig:pause_summary}(a)].
The maximum in the refined scan occurs at $s_p=0.40$, where $P_G=0.862$, approximately twice the standard-anneal baseline.
At the same pause location, the mean decoded logical energy decreases from $-78.36$ for the baseline to $-80.88$.
The coincidence of the enhanced ground-state population with the lower decoded energy identifies a pronounced relaxation window in the intermediate part of the anneal.

The improvement in optimization is not accompanied by uniform sampling of the ground-state manifold.
At $s_p=0.40$, the state-level distance is $d_{\rm TV}^{\rm st}=0.308$, compared with the baseline value $0.231$ [Fig.~\ref{fig:pause_summary}(b)].
The corresponding sector-level value is $d_{\rm TV}^{\rm sec}=0.161$, compared with the baseline value $0.156$ [Fig.~\ref{fig:pause_summary}(c)].
More generally, neither sampling metric exhibits an improvement that tracks the pronounced maximum of $P_G$ around $s_p\simeq0.4$.

Figure~\ref{fig:pause_summary}(d) resolves the same behavior at the level of individual defect sectors.
Each row gives the conditional probability of the $83$ defect sectors after restricting the samples to the exact ground-state manifold.
For a uniform sector distribution, every sector would have probability $1/83\simeq0.012$, or approximately $1.20\%$.
Instead, the heat map displays persistent over- and under-representation of particular sectors throughout the refined pause window, including near $s_p=0.40$ where $P_G$ is maximal.
The pause therefore strongly enhances relaxation into the ground-state manifold without producing uniform mixing within that manifold.

\section{Discussion}
\label{sec_dis}

The present benchmark makes it possible to distinguish optimization success from sampling within an exactly known degenerate ground-state manifold. 
The two quantities are clearly not equivalent. 
In particular, the QPU can reach a large fraction of the exact ground-state manifold while retaining a substantial deviation from the uniform distribution. 
Conversely, a protocol with a smaller ground-state hit probability can exhibit a smaller conditional sampling bias once a ground state is reached. 
The ground-state success probability $P_G$ alone therefore does not fully characterize the behavior of the annealer on a degenerate optimization problem.

The annealing-time and pause dependences give additional information about the dynamics. 
For the larger frustrated instances, fast annealing produces very small ground-state populations, whereas $P_G$ increases strongly with standard annealing time. 
More directly, for the frustrated odd-$N$ instance studied in detail, inserting a $100~\mu{\rm s}$ pause produces a pronounced enhancement of $P_G$ in an intermediate window around $s_p\simeq0.4$. 
Pauses introduced earlier or later are less effective. 
These observations support a thermally assisted quantum-annealing picture: transverse-field fluctuations remain relevant in the intermediate part of the anneal,
while coupling to the environment enables relaxation toward low-energy configurations.

An important feature of the pause experiment is that the enhanced relaxation into the ground-state manifold is not accompanied by uniform mixing within that manifold. 
Near the pause location where $P_G$ is maximal, neither $d_{\rm TV}^{\rm st}$ nor $d_{\rm TV}^{\rm sec}$ shows a corresponding suppression, and the sector-resolved distribution remains visibly nonuniform. 
Thus, relaxation into the ground-state manifold and redistribution among degenerate ground states represent distinct aspects of the annealing dynamics. 
The pause primarily improves the former without eliminating the bias associated with the latter.

The microscopic origin of the residual sampling bias cannot be identified uniquely from the present data. 
Possible contributions include freeze-out during the late stage of the anneal, analog-control errors, and details of the physical embedding. 
The strong imbalance between the two globally spin-reversed ground states observed for the $N\equiv2\pmod4$ instance shows that device or implementation dependent symmetry breaking can be significant even when the corresponding logical states are exactly degenerate. 
At the same time, the nonuniform defect-sector distribution for the frustrated odd system shows that the observed bias is not restricted to an imbalance between global-spin-flip partners. 
Separating these different sources of bias would require additional control over gauge choice, logical relabeling, and physical embedding.

The simulated-annealing results should also be interpreted as baselines for the protocols used here rather than as a statement of general classical hardness. 
In particular, the large total-variation distances obtained for the difficult odd instance are partly a consequence of the very small number of ground-state samples generated by SA. 
The main role of the classical comparison is therefore to provide a reference for optimization and sampling under the specified conditions. 
The broader value of the model is that its exact ground-state manifold and defect structure allow optimization success, state-level sampling, and defect-sector sampling to be examined separately within the same controlled benchmark.

\section{Conclusion}
\label{sec_con}

We have used an exactly characterized Ising ring with maximally separated interactions as a benchmark for open-system quantum annealing. 
The exact ground-state classification allows optimization success and sampling within a degenerate ground-state manifold to be evaluated separately. 
We find that the QPU can reach the ground-state manifold with high probability while sampling the degenerate ground states nonuniformly, demonstrating that optimization success does not imply fair sampling.

The annealing-time dependence and pause experiments further show that relaxation into the ground-state manifold is strongly enhanced during an intermediate stage of the anneal, supporting a thermally assisted quantum-annealing picture. 
This enhanced relaxation does not remove the residual state- and defect-sector sampling bias. 
The model therefore provides a controlled benchmark for distinguishing ground-state optimization from intramanifold sampling in open-system quantum annealing.

\begin{acknowledgments}
  We thank Paul A. Warburton and Natasha Feinstein for fruitful discussions.
  M.H. acknowledges financial support from the Advanced Institute for AI So-Go-Chi Informatics, Tohoku University, through its FY2026 Research Grant.
  M.O. acknowledges financial support from the Cross-ministerial Strategic Innovation Promotion Program (SIP) of the Cabinet Office (No. 23836436).
\end{acknowledgments}

\appendix

\section{Derivation of the exact ground-state classification}
\label{app_gs}

This appendix derives the ground-state energies and degeneracies collected in Table~\ref{tab:ground_state_summary}.
The argument is elementary and proceeds in three steps.
We first pass from spin variables to nearest-neighbor bond variables, which converts the Hamiltonian into a counting functional subject to a single global parity constraint.
We then express each maximally separated correlation as a product of consecutive bond variables, so that the maximally separated interactions impose a local parity rule on windows of the ring.
Minimizing the resulting functional in each congruence class of $N$ modulo four gives the classification.
Throughout we use the antiferromagnetic couplings of Eq.~\eqref{eq:benchmark_couplings} and assume $N\ge5$.

\subsection{Bond variables and the global parity constraint}

We introduce the nearest-neighbor bond variables
\begin{align}
  \tau_i = \sigma_i\sigma_{i+1},
  \qquad i\in\mathbb{Z}_N ,
  \label{eq:app-bond}
\end{align}
which obey the single global constraint
\begin{align}
  \prod_{i=0}^{N-1}\tau_i = 1 .
  \label{eq:app-constraint}
\end{align}
Conversely, every assignment $\{\tau_i\}$ satisfying Eq.~\eqref{eq:app-constraint} is realized by exactly two spin configurations, related by a global spin reversal.
A nearest-neighbor bond is satisfied by the antiferromagnetic coupling when $\tau_i=-1$, and we call a bond with $\tau_i=+1$ a \emph{defect}.
Writing $D_1$ for the number of defects, Eq.~\eqref{eq:app-constraint} is equivalent to the parity condition
\begin{align}
  D_1 \equiv N \pmod 2 .
  \label{eq:app-parity}
\end{align}
This is the only constraint that the ring geometry imposes on the defect positions.

Let $u$ denote the number of unsatisfied maximally separated bonds.
Using $|J_1|=1$ and $|J_*||E_*|=N/2$, which holds in both parity sectors by Eq.~\eqref{eq:benchmark_couplings}, the energy of an arbitrary configuration is
\begin{align}
  H_N = -\frac{3N}{2} + C ,
  \qquad
  C \equiv 2D_1 + 2|J_*|\,u ,
  \label{eq:app-cost}
\end{align}
so that $2|J_*|=2$ for even $N$ and $2|J_*|=1$ for odd $N$.
Since $C\ge0$, the ground states are exactly the configurations minimizing the cost $C$, and the problem is reduced to a combinatorial minimization over defect patterns.

\subsection{Window rule for the maximally separated bonds}

Let $\ell$ denote the step length of the maximally separated edges, so that $\ell=N/2$ for even $N$ and $\ell=m=(N-1)/2$ for odd $N$.
Each such bond connects $\sigma_i$ and $\sigma_{i+\ell}$, and telescoping Eq.~\eqref{eq:app-bond} along the ring gives
\begin{align}
  \sigma_i\sigma_{i+\ell}
  = \prod_{k=i}^{i+\ell-1}\tau_k
  = (-1)^{\ell-n_i} ,
  \label{eq:app-window}
\end{align}
where
\begin{align}
  n_i = \#\{\,k\in W_i : \tau_k=+1\,\},
  \qquad
  W_i = \{i,i+1,\dots,i+\ell-1\} ,
  \label{eq:app-nwin}
\end{align}
is the number of defects contained in the window $W_i$ of $\ell$ consecutive bonds.
A maximally separated bond is therefore satisfied if and only if
\begin{align}
  n_i \not\equiv \ell \pmod 2 .
  \label{eq:app-rule}
\end{align}
Equation~\eqref{eq:app-rule} is the origin of the dependence on $N$ modulo four: the parity of $\ell$ is fixed by $N\bmod4$, and it determines whether the maximally separated bonds require an even or an odd number of defects in every window.

\subsection{Even $N$}

For even $N$ we have $\ell=N/2$, $2|J_*|=2$, and $D_1$ even by Eq.~\eqref{eq:app-parity}.

Consider first $N\equiv2\pmod4$, for which $\ell$ is odd.
The defect-free configuration $D_1=0$ gives $n_i=0$ for every $i$, which satisfies Eq.~\eqref{eq:app-rule} and hence $u=0$.
This configuration attains $C=0$, the absolute minimum of Eq.~\eqref{eq:app-cost}, and it is the only one that does so.
The ground-state energy and degeneracy are therefore
\begin{align}
  E_0 = -\frac{3N}{2} ,
  \qquad
  |G| = 2 ,
  \label{eq:app-mod2}
\end{align}
the two states being the N\'eel configuration and its global spin reversal.

For $N\equiv0\pmod4$ the step length $\ell$ is even, and Eq.~\eqref{eq:app-rule} instead demands that every window contain an odd number of defects.
The defect-free configuration now leaves all maximally separated bonds unsatisfied, giving $u=N/2$ and $C=N$.
The smallest admissible nonzero defect number is $D_1=2$, and two defects at bond positions $a$ and $b$ produce $n_i\in\{0,1,2\}$.
Requiring $n_i$ to be odd for all $i$ forces $n_i=1$ identically, and comparing consecutive windows through
\begin{align}
  n_{i+1}-n_i
  = \mathbb{1}\!\left[\,i+\tfrac{N}{2}\in\{a,b\}\,\right]
  - \mathbb{1}\!\left[\,i\in\{a,b\}\,\right]
  \label{eq:app-shift}
\end{align}
shows that the defect pair must be invariant under a shift by $N/2$.
Since $a\ne b$, this fixes
\begin{align}
  b = a + \frac{N}{2} ,
  \label{eq:app-antipodal}
\end{align}
and any other placement of two defects leaves $u>0$ and hence $C>4$.
Configurations with $D_1\ge4$ obey $C\ge8$, and the defect-free branch has $C=N\ge8$ for $N\ge8$, so the minimum cost is $C=4$ and is attained only by the antipodal defect pairs of Eq.~\eqref{eq:app-antipodal}.
There are $N/2$ such pairs, each corresponding to two spin configurations, and therefore
\begin{align}
  E_0 = -\frac{3N}{2} + 4 ,
  \qquad
  |G| = N .
  \label{eq:app-mod0}
\end{align}

\subsection{Odd $N$}

For odd $N$ it is convenient to use the relabeled representation of Eq.~\eqref{eq:odd_relabelled_hamiltonian}, in which the maximally separated edges become step-one edges and the original nearest-neighbor edges become step-two edges.
We define the bond variables of the relabeled ring,
\begin{align}
  t_j = \hat\sigma_j\hat\sigma_{j+1},
  \qquad j\in\mathbb{Z}_N ,
  \qquad
  \prod_{j=0}^{N-1} t_j = 1 ,
  \label{eq:app-tvar}
\end{align}
so that $t_j$ measures a maximally separated bond while $\hat\sigma_j\hat\sigma_{j+2}=t_jt_{j+1}$ measures an original nearest-neighbor bond.
In these variables the two counting quantities entering Eq.~\eqref{eq:app-cost} read
\begin{align}
  u = \#\{\,j : t_j = +1\,\},
  \qquad
  D_1 = \#\{\,j : t_j = t_{j+1}\,\} = N - W ,
  \label{eq:app-uD}
\end{align}
where $W$ is the number of domain walls of the sequence $\{t_j\}$ on the ring.
The constraint in Eq.~\eqref{eq:app-tvar} makes $u$ odd, and $W=2b$ is even, with $b$ the number of maximal blocks of $t_j=+1$.
Excluding the uniform case $u=N$, which gives $C=3N$, the cost becomes
\begin{align}
  C = 2N - 4b + u ,
  \qquad
  1 \le b \le \min(u,\,N-u) ,
  \qquad
  u \ \text{odd} .
  \label{eq:app-costodd}
\end{align}
The upper bound on $b$ is saturated by arranging the minority species as isolated singletons, so minimizing over $b$ at fixed $u$ gives
\begin{align}
  C_{\min}(u) =
  \begin{cases}
    2N - 3u, & u \le m ,\\[2pt]
    5u - 2N, & u \ge m+1 .
  \end{cases}
  \label{eq:app-Vshape}
\end{align}
Equation~\eqref{eq:app-Vshape} is a piecewise-linear function with its minimum at $u\simeq N/2$, and the parity of $m$ decides which of the two admissible odd values of $u$ closest to $N/2$ is selected.

For $N\equiv3\pmod4$ the step length $m$ is odd, so $u=m$ is allowed and gives $C=(N+3)/2$, whereas the competing value $u=m+2$ gives $C=(N+15)/2$.
For $N\equiv1\pmod4$ the step length $m$ is even, so $u=m$ is excluded; the admissible values $u=m-1$ and $u=m+1$ give $C=(N+9)/2$ and $C=(N+5)/2$, respectively.
In both congruence classes the minimizing configuration has $b=\min(u,N-u)=m$, and therefore
\begin{align}
  D_1 = N - 2m = 1 .
  \label{eq:app-onedefect}
\end{align}
This is the analytic statement that the odd system carries exactly one unavoidable nearest-neighbor defect, as required by the odd cycle formed by the step-two edges.
Substituting the minimal costs into Eq.~\eqref{eq:app-cost} yields
\begin{align}
  E_0 =
  \begin{cases}
    \displaystyle -N + \frac{5}{2}, & N\equiv1 \pmod 4 ,\\[6pt]
    \displaystyle -N + \frac{3}{2}, & N\equiv3 \pmod 4 ,
  \end{cases}
  \label{eq:app-E0odd}
\end{align}
in agreement with Eq.~\eqref{eq:gs_odd_energy} of the main text.
The difference between the two classes originates entirely in the number of maximally separated bonds left unsatisfied by the single defect: a window of length $m$ contains the defect for exactly $m$ of the $N$ maximally separated bonds, and Eq.~\eqref{eq:app-rule} assigns the satisfied bonds to the windows containing the defect when $m$ is even and to those avoiding it when $m$ is odd, giving $u=(N+1)/2$ and $u=(N-1)/2$, respectively.

Once the defect position $a\in\mathbb{Z}_N$ is specified, all remaining nearest-neighbor bonds are antiparallel and the spin configuration is determined up to a global spin reversal.
The exact ground-state manifold therefore consists of
\begin{align}
  |G| = 2N
  \label{eq:app-degodd}
\end{align}
states, organized into $|S|=N$ defect sectors of two states each, as used in Sec.~\ref{sub_samp}.
For the instance studied in detail in the main text, $N=83\equiv3\pmod4$, Eqs.~\eqref{eq:app-E0odd} and \eqref{eq:app-degodd} give $E_0=-81.5$, $|G|=166$, and $|S|=83$.

\bibliographystyle{unsrt}
\bibliography{reference}

\end{document}